\documentclass[prl,aps,twocolumn,amsmath,amssymb,superscriptaddress]{revtex4-1}

\usepackage{graphicx}
\usepackage{amsfonts}
\usepackage{color}
\usepackage{amstext}
\usepackage{color}

\begin{document}

\title{Mesoscopic superpositions of Tonks-Girardeau states and the Bose-Fermi mapping}

\author{M. A. Garc\'{\i}a-March} 
\affiliation{Departament d'Estructura i Constituents de la Mat\`eria, 
Univ. de Barcelona, 08028 Barcelona, Spain}
\affiliation{ICFO-Institut de Ci\`encies Fot\`oniques, 
Parc Mediterrani de la Tecnologia, 08860 Barcelona, Spain}
\author{A. V. Yuste-Roca} 
\affiliation{Departament d'Estructura i Constituents de la Mat\`eria, 
Univ. de Barcelona, 08028 Barcelona, Spain}
\affiliation{F\'{\i}sica Te\`orica: Informaci\'o i Fen\`omens Qu\`antics, 
Departament de F\'{\i}sica, Universitat Aut\`onoma de Barcelona, 
08193 Bellaterra (Barcelona), Spain}
\author{B. Juli\'a-D\'iaz}
\affiliation{Departament d'Estructura i Constituents de la Mat\`eria, 
Univ. de Barcelona, 08028 Barcelona, Spain}
\affiliation{ICFO-Institut de Ci\`encies Fot\`oniques, 
Parc Mediterrani de la Tecnologia, 08860 Barcelona, Spain}
\author{A. Polls}
\affiliation{Departament d'Estructura i Constituents de la Mat\`eria, 
Univ. de Barcelona, 08028 Barcelona, Spain}

\begin{abstract}
We study a one dimensional gas of repulsively interacting ultracold 
bosons trapped in a double-well potential as the atom-atom interactions 
are tuned from zero to infinity. We concentrate on the properties of 
the excited states which evolve from the so-called NOON states to  the
NOON Tonks-Girardeau states. The relation between the latter and the 
Bose-Fermi mapping limit is explored. We state under which conditions 
NOON Tonks-Girardeau states, which are not predicted by the Bose-Fermi 
mapping, will appear in the spectrum. 
\end{abstract}

\maketitle

The recent ground-breaking  experiments on the trapping of a few fermions or 
bosons have stir up the theoretical interest in small ultracold quantum 
gases~\cite{2010XeOE,2011SerwaneScience,2012ZurnPRL,2013WenzScience,
2013BourgainPRA,2014PaganoNatphys}. 
The atoms in these experiments can be often regarded as being effectively 
trapped in one dimension and interacting through contact interactions. This 
is a system with many examples of exact solvability 
(see, e.g.,~\cite{2008GiorginiRMP,2011CazalillaRMP,2013GuanRMP}). 
A  step forward in these experiments is to consider other geometries, 
a goal which is already accomplished for double-well 
potentials~\cite{2014MurmannArxiv}.

Experiments have already reached the strongly interacting 1D gas of 
bosons~\cite{Kinoshita,Paredes}, exploring the mapping of repulsive 
bosons into a system of ideal fermions to form the so-called 
Tonks-Girardeau (TG) gas~\cite{Girardeau1960,Girardeau2001}. 
Simultaneously, double-well experiments with ultracold 
bosons~\cite{2005AlbiezPRL,2007LevyNature} should sooner than later 
allow one to produce macroscopic superpositions~\cite{1998CiracPRA}. So,  combining the two may allow one to produce macroscopic 
superpositions of strongly interacting TG gases. In this letter we settle 
under which conditions the  superpositions in the Fock regime, 
so called NOON (cat-like) states~\cite{2009WeissPRL,2009StreltsovPRA,2013FogartyPRA}, evolve into  superpositions of TG gases, 
termed NOON TG, as the interactions are tuned from small to infinity.  

\paragraph{\bf Model Hamiltonian.}

\begin{figure}
\includegraphics[width=0.85\columnwidth]{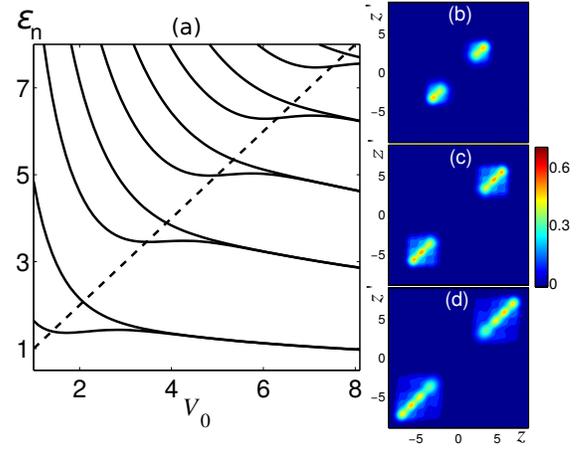}
\caption{(a) The first ten eigenvalues of the single-particle 
Hamiltonian as a function of $V_0$ are depicted. The dashed line 
represents the condition $E=V_0$. (b)-(d) One-body density matrix for 
NOON Tonks-Girardeau states of $N=2, 3$ and 4 obtained 
with $V_0=6,10$ and 12, respectively.
\label{fig1}     }
\end{figure}

Let us consider $N$ atoms trapped in a one-dimensional 
double-well potential, described by the first-quantized 
Hamiltonian 
$\hat H=\sum_{j=1}^N\left[-(1/2)\partial^2_{z_j^2} +V_{\rm{DW}}(z_j)\right] + V_{\rm int}$, 
where the interaction is taken as a contact, 
$V_{\rm int}=g \sum_{i<j} \delta(z_i-z_j)$ and where $\hbar=m=1$. 
To model the double-well potential we use a simplified 
Duffing potential,
$ V_\mathrm{DW}(z)=V_0\left(-8(z/V_0)^2 +16 (z/V_0)^4 +1\right)$. 
This Duffing potential has two minima equal to zero at 
$z=\pm V_0/2$ and a local maximum equal to $V_0$ at $z=0$. 
$V_0$ is thus the barrier height and the distance between 
the wells. In our study, we take 
$V_0\in[V_0^{\mathrm{min}}, V_0^{\mathrm{max}}]$, with $V_0^{\mathrm{min}}$ 
a small number and   $V_0^{\mathrm{max}}$ large enough to get a  
degeneracy of the order of $10^{-10}$ between the first two 
single-particle eigenstates (that is  $V_0^{\mathrm{max}}\simeq 12$). 
By increasing $V_0$, the confining potential goes from a quartic, for 
$V_0\to 0$, into increasingly deeper double-well potentials. 
Let $\phi_{n}$ be the $n$-th eigenstate of the single-particle 
Hamiltonian $H _{\mathrm{sp}}= (1/2) \partial^2_{z^2} +V_\mathrm{DW}(z)$ 
with energy $\varepsilon_n$. As $V_0$ is increased, quasi-degeneracies 
between the single-particle energies are introduced, see 
Fig.~\ref{fig1}(a). These occur for values of the corresponding 
quasi-degenerate pair of energies smaller than $V_0$. 

\begin{figure}[t]
\includegraphics[width=0.95\columnwidth]{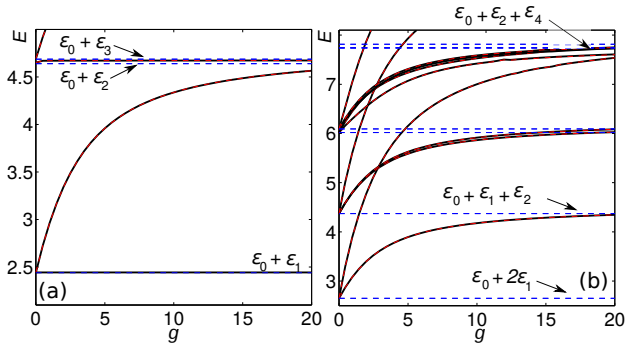}
\caption{Energy spectra for $N=2$ (a) and $N=3$ (b) with $V_0=5$ 
and $V_0=10$, respectively. The blue dashed lines in both panels 
represent the energies expected in the Fermionized limit, 
$g\to\infty$ (we explicitly label a few of them to help to 
understand this limit). These energies have been calculated by extrapolating the results 
obtained with  an increasing number of modes, with a maximun number of  modes of  30 and 24 for $N=2$ and $N=3$, respectively.
 \label{fig6}}
\end{figure}

\paragraph{\bf Spinless fermions.} For $N$ spinless fermionic atoms, 
the ground and all excited states $\Psi^{\rm f}_{k}$ of $\hat{H}$ are Slater 
determinants built 
up from $N$ single-particle eigenstates $\phi_{n}$. This wave 
functions fulfil the Pauli exclusion principle, i.e. are zero 
whenever $z_i=z_j$, and therefore do not feel the contact 
interaction potential. Also, these solutions have a sign change 
at these points, i.e.  for fermions, the wavefunction has to 
change sign when  an atom of coordinate $z_i$ is interchanged with 
one at $z_j$. The energies of the many-body eigenstates $\Psi^{\rm f}_{k}$ are the summation of the energies
of the single-particle states used to build the corresponding Slater determinant.

For $N=2$ fermions, which will be used as a limiting case 
in the forthcoming discussions, the ground state is 
$\Psi^{\rm f}_{0}(z_1,z_2) = 1/\sqrt{2} \,{\rm Slater}(\phi_0,\phi_1)
= 1/\sqrt{2}\left(
\phi_0(z_1)\phi_1(z_2)-\phi_1(z_1)\phi_0(z_2) \right)
$ 
and has energy $E_0=\varepsilon_0+\varepsilon_1$ for all $V_0$. 
For large $V_0$, the first two single-particle states are 
quasi-degenerate, $\varepsilon_1=\varepsilon_0+\varepsilon_{01}$, 
with $\varepsilon_{01}\ll \varepsilon_0$. The next four excited 
states are $\Psi^{\rm f}_{1}\sim{\rm Slater}(\phi_0,\phi_2)$, 
$\Psi^{\rm f}_{2}\sim{\rm Slater}(\phi_1,\phi_2)$, $\Psi^{\rm f}_{3}\sim{\rm Slater}(\phi_0,\phi_3)$, 
and $\Psi^{\rm f}_{4}\sim{\rm Slater}(\phi_1,\phi_3)$. For $V_0\gtrsim4.5$, 
the second and third single-particle states are also quasi-degenerate 
(see Fig.~\ref{fig1}), and therefore their energies can be written as 
$\varepsilon_3=\varepsilon_2+\varepsilon_{23}$, with 
$\varepsilon_{23} \ll \varepsilon_2$. Thus, for $V_0 \gtrsim 4.5$ the first 
four excited states are quasidegenerate and have energies 
$E_1=\varepsilon_0+\varepsilon_2$, 
$E_2=\varepsilon_0+\varepsilon_2+\varepsilon_{01}$, 
$E_3=\varepsilon_0+\varepsilon_2+\varepsilon_{23}$, and 
$E_4=\varepsilon_0+\varepsilon_2+\varepsilon_{01}+\varepsilon_{23}$. 
For two atoms, the next excited state has a finite energy gap with 
this manifold (see gap between $\varepsilon_3$ and $\varepsilon_4$ 
for large $V_0$ in Fig.~\ref{fig1}). 

For sufficiently large $V_0$, such that a sizeable amount of pairs of single-particle states are quasidegenerate, we can define the following 
single-particle wavefunctions, 
\begin{equation}
\label{eq:locbasis}
  \phi_n^{j} = \left\{\def\arraystretch{1.2}%
  \begin{array}{@{}c@{\quad}l@{}}
    \phi_{n}^L=\dfrac{1}{\sqrt{2}}(\phi_{2n}+\phi_{2n+1})  &\\
    \phi_{n}^R=\dfrac{1}{\sqrt{2}}(\phi_{2n}-\phi_{2n+1}) & \,, n=0, 1, \dots  \,.\\
  \end{array}\right. 
\end{equation}
Here $j=L(R)$ stands for left (right). These $\phi_n^{j}$ functions 
are mostly localized either in the left or right well when the pair 
of delocalized functions $\phi_{n}$ used to construct them are 
quasi-degenerate. For large interaction energies, the small degeneracy 
breaking between the states will in some cases become irrelevant, and 
a much simpler picture is obtained in the localized basis.

\paragraph{\bf Interacting bosons in a double well.} Using the 
single-particle eigenfunctions to build the Fock basis, 
the many-body Hamiltonian for bosons reads
\begin{align}
\mathcal{\hat H}=\sum_{i=1}^{M}\varepsilon_{i}\hat a_i^{\dagger}\hat a_i
+\dfrac{g}{2}\sum_{k,l,m.n}^{M}I_{k,l,m,n}\hat a_k^{\dagger}\hat a_l^{\dagger}\hat a_m\hat a_n
\label{ham},
 \end{align}
with $I_{k,l,m,n}=\int dz\phi_k(z)\phi_l(z)\phi_m(z)\phi_n(z)$, where 
$[\hat a_{k}^{\dagger},\hat a_{i}]=\delta_{ik}$ and 
$[\hat a_{k}^{\dagger},\hat a_{i}^{\dagger}]=\left[\hat a_{k},\hat a_{i}\right]=0$.
A Fock vector is written as 
$|n_0,n_1,\dots n_M\rangle={\cal N}(a^\dagger_0)^{n_0}\dots,(a^\dagger_M)^{n_M}|{\rm vac}\rangle$, 
where $|{\rm vac}\rangle$ is the vacuum, $n_i$ is the number of atoms in 
mode $i$ and ${\cal N}^{-1}=\sqrt{n_0! \dots n_M!}$.

\begin{figure}
\includegraphics[width=0.95\columnwidth]{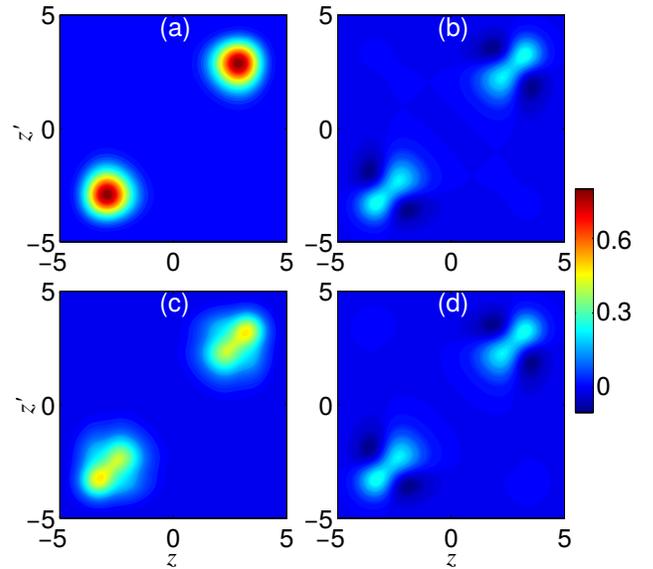}
\caption{One-body density matrices of two of the four excited 
states for two bosons in a double-well for $V_0=5$. The upper (lower) 
row is computed with $g=0.5$ ($g=20$). One-body density matrices for 
a NOON state, panel (a), and its evolved state, panel (c), and a 
non-interacting state, panel (b) and its evolved state, panel (d). 
\label{fig5}}
\end{figure}

\paragraph{\bf Beyond the bimodal Fock regime.} For small 
interactions the system remains bimodal, allowing to explore with a two-mode model from the Josephson to the Fock regimes\cite{2001LeggetRMP,2012DaltonJMO}.
As interactions are increased, the system approaches the strongly 
correlated regime~\cite{2004StreltsovPRA,2008MurphyPRA,2008ErnstPRA}. Let us explore in detail 
this transition, using as driving examples the $N=2$ and $N=3$ cases.

Assuming a barrier high enough to have two quasi-degenerate pairs 
in the lower part of the single-particle spectrum, the interaction 
energy will make the energy gaps irrelevant. In this case, the 
simplest picture is provided using the {\it localized} single-particle 
basis defined in Eq.~(\ref{eq:locbasis}). Using this basis, the Fock 
vectors can be written as, 
$|n_0^L,n_0^R, n_1^L,n_1^R, \dots\rangle_{\rm lo} = {\cal N} 
(a^{L \, \dagger}_0)^{n_0^L}(a^{R\, \dagger}_0)^{n_0^R}\dots
(a^{L \, \dagger}_{M/2})^{n_{M/2}^L} (a^{R\, \dagger}_{M/2})^{n_{M/2}^R} |{\rm vac}\rangle$ with 
${\cal N}^{-1}= \sqrt{n_0^R! n_0^L! \ \dots n_{M/2}^R! n_{M/2}^L!} $.

For instance, in the $N=2$ case, once the interaction is larger than 
the gap, the ground state is well approximated by $|1,1,0,\dots \rangle_{\rm lo}$. 
This state has one atom in each well and is thus unaffected by the interaction. 
Its energy remains constant as $g$ is increased, see Fig.~\ref{fig6}(a). 
The first two excited states, quasidegenerated with the ground state in the 
non-interacting case, are the NOON states, 
$|{\rm NOON}_2\rangle=(|0,2,0,\dots\rangle_{\rm lo} \pm |2,0,0,\dots,\rangle_{\rm lo})/\sqrt{2}$, 
which contain a delocalized pair of interacting atoms (two atoms in the same well). Their energy grows linearly 
with $g$ for small $g$, see Fig.~\ref{fig6}(a). The next excited states 
involve more than two modes and are clearly gapped in absence of 
interactions~\cite{2007DounasFrazerPRL,2012GarciaMarchFiP}. They read, 
$(|0,1,1,0,\dots\rangle_{\rm lo} \pm |1,0,0,1,0,\dots\rangle_{\rm lo})/\sqrt{2}$, 
and $(|0,1,0,1,0,\dots\rangle_{\rm lo} \pm |1,0,1,0,\dots\rangle_{\rm lo})/\sqrt{2}$. 
As before, the first two are mostly non-interacting and thus their 
energies remain constant as $g$ is increased. The latter, however, 
have one pair of interacting atoms, 
their energy grows linearly with similar slope as the 
$|{\rm NOON}_2\rangle$ states, see Fig.~\ref{fig6}(a). 

\begin{figure}
\includegraphics[width=1.\columnwidth]{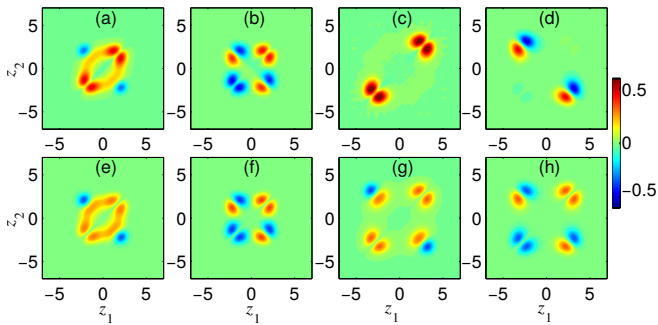}
\caption{Comparison of two of the four excited numerical wave 
functions for $g=20$ (upper row) and the analytic ones 
$\Psi^{\rm b}_1$ (e,g) and $\Psi^{\rm b}_3$ (f,h) obtained through 
the Bose-Fermi mapping (lower row). Panels (a), (b), (e) and (f) are obtained 
with $V_0=4$, while panels (c), (d), (g) and (h) correspond to $V_0=5$. $N=2$.
\label{figapp}}
\end{figure}

For $N>2$ there are no noninteracting states. For $N=3$ the lowest energy 
manifold at small interactions is 
$( |1,2,0,\dots\rangle_{\rm lo} \pm |2,1,0,\dots,\rangle_{\rm lo})$
and the NOON states, $|{\rm NOON}_3\rangle=( |3,0,0,\dots\rangle_{\rm lo} 
\pm |0,3,0,\dots,\rangle_{\rm lo})$. The pair of atoms on the same well 
in the first ones are responsible for the linear increase in the 
energy as a function of $g$, with same slope as for $N=2$ discussed above. 
In contrast, the NOON states have 3 interacting pairs, with a 
correspondingly larger slope, see Fig.~\ref{fig6}(b). 

In the general case, the lowest energy manifold at small $g$ is 
$|N,0,0, \dots\rangle_{\rm lo} \pm |0,N,0,\dots\rangle_{\rm lo}$, 
$|1,N-1,0\dots\rangle_{\rm lo} \pm |N-1,1,0\dots\rangle_{\rm lo}$, etc. 
Our interest at this point is to disentangle the fate 
of these states as the interaction is varied from the Fock regime 
into the strongly interacting regime. In particular, what happens 
with the states in which more than one atom populates each well 
in the Fock regime. The driving intuition is twofold. On one hand, 
we know that for bosons in a single well, the system evolves into 
a TG gas, i.e. the NOON would directly evolve into a 
NOON TG gas. On the other hand, the Bose-Fermi mapping 
theorem can be directly applied to the system, providing exact 
solutions in the infinitely interacting case, which will be shown to 
be in contradiction with the first one.

\paragraph{\bf From NOON to the NOON Tonks-Girardeau.} The $N=2$ and 
$N=3$ cases are again very illustrative. We note that 
energetically the fate of the NOON states in $N=2$ is very similar 
to what happens to the 
$(|1,2,0,\dots\rangle_{\rm lo} \pm |2,1,0,\dots\rangle_{\rm lo})/\sqrt{2}$ 
states in $N=3$, see Fig.~\ref{fig6}, i.e. the third particle in the $N=3$ 
plays an spectator role as $g$ is increased. The latter 
takes place also in the general $N>2$ case, where none of the 
eigenstates is non-interacting and in which upon increasing $g$ the 
system is expected to increase correlations to avoid the interaction. 
Thus, let us first disentangle the fate of the NOON states at $N=2$. 
From Fig.~\ref{fig6}(a) we see that as $g$ is further increased, 
their energies saturate to a constant value which actually approaches 
$\varepsilon_0+\varepsilon_2$. This is also essentially the energy of 
the states evolved from  
$(|0,1,1,0,\dots\rangle_{\rm lo} \pm |1,0,0,1,0,\dots\rangle_{\rm lo})/\sqrt{2}$, 
which are non-interacting. Thus, in the strongly interacting regime, 
there are four  quasi-degenerate states.  

The evolution with increasing $g$ of their one-body density matrices (OBDM)
is very telling. The NOON states, evolve from the $|{\rm NOON}_2\rangle$ 
states at $g=0.5$, see Fig.~\ref{fig5}(a), to distributions with 
two peaks per well (or three and four peaks per well for $N=3$ and 
$N=4$, respectively, see Fig.~\ref{fig1}(c,d)), resembling the OBDM 
for two TG gases, Fig.~\ref{fig5}(c). 
The OBDM for the non-interacting states similarly  shows two-peaks per well, as would correspond to populating the 
single-particle states $\phi_{2,3}$, which have one node in each well. 

Let us now examine the prediction from the Bose-Fermi mapping theorem. 
The rigorous Bose-Fermi mapping theorem is established by noting that 
the bosonic problem in the $g\to\infty$ is equivalent to the fermionic 
problem when imposing the boundary condition that the wave functions have to 
vanish if $z_j=z_{j'}$. This boundary condition is obeyed by the fermions 
for free because it is imposed by the Pauli exclusion principle. Then, 
the bosonic wave functions $\Psi^{\rm{b}}$ can be obtained from the 
fermionic ones $\Psi^{\rm{f}}$ after symmetrizing them, 
$\Psi^{\rm{b}}=A\Psi^{\rm{f}}\,\,\, \mbox{with}\,\, 
A=\prod_{j>j'}\mbox{sgn}(z_j-z_{j'})$, 
with $\mbox{sgn}(z)$ the sign function. The analytic form of the 
wave functions for the case of fermions was discussed above. The 
Bose-Fermi mapping theorem applies equally regardless of the 
value of $V_0$. 

In Fig.~\ref{figapp} we compare the predictions of the 
Bose-Fermi mapping, to the actual numerical results, for $N=2$ and 
for $V_0=4$ and $V_0\gtrsim4.5$. For $V_0=4$, the 
Bose-Fermi mapping, panels (e,f), provides an accurate representation 
of the numerically obtained results, panels (a,b). Both the NOON and 
the non-interacting states, evolve at 
$g=20$ to the states predicted by the Bose-Fermi mapping. Note that 
in this case, the barrier is not high enough to substantially localize 
the single-particle ground and first excited state, see Fig.~\ref{fig1}. 
None of the many-body states is however a NOON TG state. 

\begin{figure}
\includegraphics[width=1.\columnwidth]{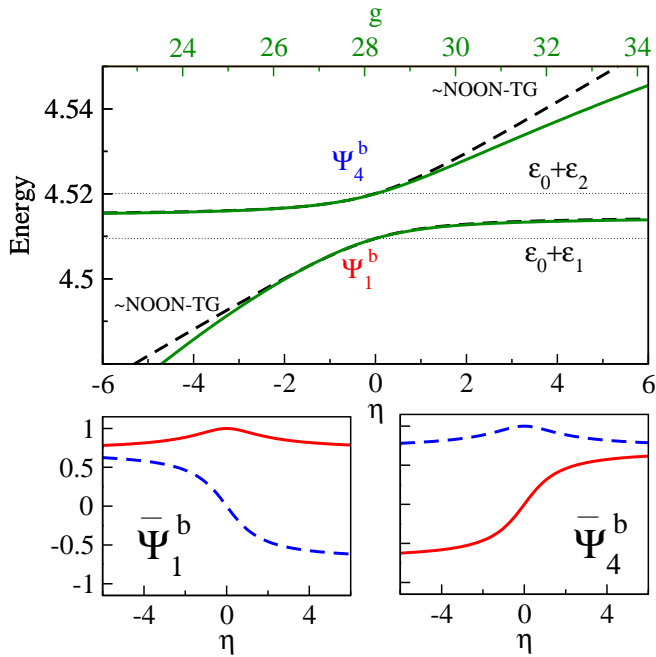}
\caption{Detailed analysis of the $g\to\infty$ limit for $V_0=5.5$ 
(upper panel). Two of the first excited states for $N=2$ and $V_0$, 
computed numerically, solid lines, 
compared to the results of the perturbation theory when the width of the perturbation is $\sigma\simeq0.07$ (see the text for further details), dashed lines.  The small panels depict the composition of the dressed states, 
$\bar{\Psi}_1^b$ and $\bar{\Psi}_4^b$: $\langle \bar{\Psi}_k^b |\Psi_1^b \rangle$ (solid, red) and  $\langle \bar{\Psi}_k^b |\Psi_4^b \rangle$ (dashed, blue). 
A similar qualitative picture is obtained for states $\bar{\Psi}_2^b$ and $\bar{\Psi}_3^b$. 
\label{fig55}}
\end{figure}

A surprising result appears however as we increase the barrier height, 
going from $V_0=4$ to $V_0=5$ in Fig.~\ref{figapp}. In this case, the 
prediction of the Bose-Fermi mapping, panels (g,h), and the numerical 
results, panels (c,d), clearly disagree. The numerical results show less 
left-right coherence than the Bose-Fermi prediction. 
That is, there is essentially zero probability of finding two bosons in the same well, as seen in Fig.~\ref{figapp} (d), while in the Bose-Fermi mapping, the
probability is clearly non-negligible in all four states. 
It is worth mentioning that these differences, which involve 
pair-correlations, are not reflected in the density profiles, 
i.e., the diagonal of the one-body density matrices, 
Figs.~\ref{fig1}(b,c,d) and \ref{fig5} (c,d). What causes such 
discrepancy? And, more important, what would an experiment 
find? Let us note, that in all cases, the left-right symmetry is fully 
respected, i.e. it is not a consequence of spurious numerical biases 
in the numerics.

\paragraph{\bf Finite-range effects.} 
The Bose-Fermi mapping applies for delta interaction potentials 
in the strict infinite interaction case. The Slater determinant 
ensures that the atoms do not interact, but, any finite size 
correction to the delta contact potential immediately implies 
a non-zero interaction between the two atoms. In addition, any 
numerical approach has an inherent minimal distance resolved, in 
our case related to the maximun number of modes considered. Thus, 
for any number of modes, if $g$ is sufficiently large, finite 
size corrections to the contact should show up. This energy stemming 
from the finite range of the atom-atom interaction, can be 
comparable to the splitting between the four excited states 
described above, resulting in a mixture of the four states. 

This hypothesis can be tested quantitatively. Let us consider 
a finite range perturbation on the four states predicted by 
the Bose-Fermi mapping, e.g. 
$V_{\rm pert}(z_j-z_{j'}) = \eta/\sqrt{2\pi \sigma^2} 
\exp\left[-(z_j-z_{j'})^2/( 2\sigma^2)\right]$, and diagonalize 
it in the restricted space spanned by 
$\{\Psi_1^{\rm b}, \Psi_2^{\rm b}, \Psi_3^{\rm b}, \Psi_4^{\rm b}\}$. Then we obtain the dressed 
states $\{\bar{\Psi}_1^{\rm b}, \bar{\Psi}_2^{\rm b}, \bar{\Psi}_3^{\rm b}, \bar{\Psi}_4^{\rm b}\}$. 
The perturbation theory prediction can be compared to the numerical results by noting that the $\eta=0$ corresponds to the value of
$g$ for which the bose-fermi energies are reproduced
for the number of modes used. As seen in Fig.~\ref{fig55} 
the model predicts a nice avoided crossing in agreement 
with the numerics. For $\eta<0$, thus removing repulsion from the 
$g\to \infty$ limit, we have the NOON TG as the lowest states 
of the manifold, 
$\bar{\Psi}^{\rm b}_1\simeq (\Psi^{\rm b}_1+\Psi^{\rm b}_4)/\sqrt{2}$ 
(see Fig.~\ref{fig55}) and 
$\bar{\Psi}^{\rm b}_2\simeq(\Psi^{\rm b}_2+\Psi^{\rm b}_3)/\sqrt{2}$. 
For $\eta>0$, the repulsive finite size correction sets the 
non-interacting states 
$\bar{\Psi}^{\rm b}_1\simeq(\Psi^{\rm b}_1-\Psi^{\rm b}_4)/\sqrt{2}$ and 
$\bar{\Psi}^{\rm b}_2\simeq (\Psi^{\rm b}_2-\Psi^{\rm b}_3)/\sqrt{2}$
as the lowest in the manifold. 

These results, checked for $N=2,3$ and $4$ are expected to remain valid 
for larger number of atoms. NOON states will evolve to NOON Tonks-Girardeau 
states provided two conditions are met: 1) the potential barrier has to 
be large enough to ensure the existence of $N$ quasi-degenerate doublets 
in the single-particle spectrum, and 2) the actual range of the contact 
interaction is non-zero but finite,  which is what actually happens in an 
experiment, such that in the infinitely interacting limit the 
residual interaction mixes the corresponding states predicted by the 
Bose-Fermi mapping to produce NOON Tonks-Girardeau states. These 
findings, which require two-body correlations to be explicitly 
measured, will be of relevance in future experimental efforts to 
build strongly correlated states with inter-site spatial entanglement. 
As an outlook, we 
foresee that, for a general trapping potential with an associated 
single-particle eigenspectra showing quasidegeneracies, the effect 
of the finite range of the contact interactions may involve discrepancies 
between the result predicted by the Bose-Fermi mapping and the actual solution. 

We acknowledge useful discussions with Prof. Thomas Busch.  
We acknowledge  partial financial support from the 
DGI (Spain) Grant No. FIS2011-25275, FIS2011-24154 
and the Generalitat de Catalunya Grant No. 2014SGR-401. 
BJD is supported by the Ram\'on y Cajal program, MEC (Spain).
MAGM acknowledges support from  ERC Advanced
Grant OSYRIS, EU IP SIQS, EU STREP EQuaM, John Templeton Foundation, and Spanish Ministry Project FOQUS.

\end{document}